\documentclass[12pt]{iopart}
      \newcommand{\beq}{\begin{equation}}
      \newcommand{\eeq}{\end{equation}}
      \newcommand{\beqa}{\begin{eqnarray}}
      \newcommand{\eeqa}{\end{eqnarray}}
      
      \newcommand{\nn}{\nonumber}

      \newcommand{\bra}{\left\langle}
      \newcommand{\ket}{\right\rangle}
      \newcommand{\ch}{\, {{\rm ch} \, }}
       \newcommand{\sh}{\, {{\rm sh} \, }}
       \newcommand{\tah}{\, {{\rm th} \, }}
       \newcommand{\del}{\partial}

      \newcommand{\bsi}{{\bm{\sigma}}}

      \newcommand{\be}{\beta}
      \newcommand{\ga}{\gamma}
      \newcommand{\de}{\delta}

      \newcommand{\si}{\sigma}

      \newcommand{\bvphi}{\mbox{\boldmath $\varphi$}}

     \newcommand{\bh}{\mbox{\boldmath $h$}}

    \renewcommand{\(}{\left(}
    \renewcommand{\)}{\right)}

\newcommand{\rnd}[1]{\left[ #1 \right]_{\rm av}}

\usepackage{graphicx}
\usepackage{dcolumn}
\usepackage{bm}
\usepackage[dvipdfmx]{hyperref}
\usepackage{subcaption}
\usepackage{cite}

\begin{document}
\title[Non-differentiability of the effective potential and the RSB]
{Non-differentiability of the effective potential and the replica symmetry breaking 
in the random energy model
    }
 
 \author{Hisamitsu Mukaida}
 \address{Department of Physics, Saitama Medical University, 
 38 Moro-Hongo, Iruma-gun, Saitama, 350-0495, Japan
 }
\ead{mukaida@saitama-med.ac.jp}
\begin{indented}
\item[]September 2015
\end{indented}

\begin{abstract}
The effective potential for the two-replica system of the random energy model is exactly derived. 
It is an analytic function of the magnetizations of two replicas, 
$\varphi^1$ and $\varphi^2$ in the high-temperature phase.
In the low-temperature phase, where the replica symmetry breaking takes place, 
the effective potential 
becomes non-analytic when $\varphi^1 = \varphi^2$. The non-analyticity is considered as 
a consequence of the condensation of the Boltzmann measure, which is 
a typical property of a glass phase. 
\end{abstract}

 
\vspace{2pc}
\submitto{\jpa}

\maketitle

\section{Introduction}
A technical difficulty in theoretical study of quenched disordered systems 
originates from inhomogeneity due to disordered environment.  
In those systems, we first take the thermal average of physical quantities in a fixed disordered 
environment and then take the average over the disorder. 
However, if we can first average out  
the disorder, the systems become homogeneous and 
problems will be more tractable.   
Several methods to make it possible were developed in the last four decades. 
 
One of the standard method will be the replica trick \cite{MPV87,N01,D05}. 
Namely, a partition function of identical $n$ copies (replicas) of a disordered  
system is introduced and then the average over the disorder is taken.  
The resultant partition function defines a homogeneous ``replicated'' system. 
According to the replica trick, in order to extract disorder-averaged physical quantities  
from the replicated system, the zero-replica limit $n \to 0$ is taken  
despite that  $n$ is a positive integer.
Although there are several studies for exact replica approach to specific models \cite{K02, K05, OK07,D11}, 
general mathematical foundation has not been found yet \cite{HP79, D12}.

 In mean-field models such as 
the Sherrington-Kirkpatrick model \cite{SK75}
 or the random energy model (REM) \cite{D80,D81}, glassy behaviour comes out 
 together with the replica symmetry breaking (RSB).  The RSB originally means that 
the symmetry under permutation of the replica indices is (spontaneously) broken in a  
replicated system.  
It is brought about by dominance of saddle points that break the replica symmetry  
when evaluating the partition function of a replicated system.  Since the evaluation is carried out 
in the zero-replica limit $n \to 0$,  the original definition of the RSB is mathematically ambiguous.  
However, physical insights clarify that the RSB is a consequence of  contribution from metastable states,  
which can be measured by the probability distribution of the two-replica overlap.  Thus 
a well-defined order parameter of the RSB is extracted from the probability distribution,
 which is referred to as the Parisi order parameter \cite{P83,MPSTV84}. 

As for short-ranged models, Le Doussal and Wiese showed, in study of random elastic models, 
that the RSB and non-analyticity of the effective potential in the replicated system  
 appear  at the same time when the system goes into a glass phase 
from the high-temperature phase \cite{LW02,LW03}.  
If this phenomenon is confirmed in various 
quenched random systems, the non-analyticity in effective potential may be regarded as   
an indication of the RSB. 
For this reason, it is worthwhile to examine  universality of relationship 
between the non-analytic effective potential and the RSB. 

In this paper, we compute the effective potential for the replicated system consisting 
of the REM and attempt better understanding of the relationship. 
The model is simple, so that we can exactly calculate the effective potential without use of the replica trick. 
Hence, we can examine analyticity of the  effective potential without suffering from artifact by approximation and from mathematical ambiguity caused by the replica trick.  

This paper is organized as follows: in the next section, we introduce two definitions of 
the effective potential: one is defined from the Legendre transform of 
the cumulant generating function, which is adopted by the literature \cite{LW02,LW03,TT04,WL07,LMW08,TT08,TT08-2}.  
The other is so-called the constraint effective potential \cite{FK75,OWY86},   
which is defined as the free energy with an order parameter fixed.
The relationship of the two effective potentials is known, which is also described \cite{OWY86,T09}. 
In \sref{sec_REM}, we introduce the REM and compute the generating function of the replicated system with two replicas.
 In \sref{sec_eff}, the effective potential is derived by the 
Legendre transform of the generating function. The constraint effective potential is 
also computed in \sref{sec_constraint}. We discuss the origin of non-analyticity of the 
effective potential in the last section.

\section{The effective potential in a replicated system}
\label{sec_EP}
In this section,  we first recall the effective potential in a replicated system 
introduced in \cite{LW02,LW03,TT04,WL07,LMW08,TT08,TT08-2} with a little modification along 
the present work.  Next, we introduce 
 the constraint effective potential \cite{FK75,OWY86} in a replicated system.  

Consider a field theory on a lattice described by a Hamiltonian 
$H_{\rm DO}[u]$.  Here $u := \{u_i\}_i$ denotes a field variable 
with the site index $i \in \{1, ..., N \}$. Note that the Hamiltonian depends on not only 
$u$ but also disordered environment.
Suppose that $u$ is coupled  to a uniform external source $h$. The theory 
in the inverse temperature $\beta$ is described by the partition function   
$$
  Z(h) := \int {\cal D} u \  \rme^{ \be (- H_{\rm DO}[u] + N h \, \tilde{u})}, 
$$ 
where 
$$
  \tilde{u} := \frac{1}{N} \, \sum_i u_i.  
$$
When $u$ is a spin variable, $\tilde u$ corresponds with the magnetization per site.  
The ``thermal" cumulants of  $\tilde u$ (i.e., cumulants of $\tilde u$ with respect to the 
thermal average) at $h=0$ with  fixed disorder can be obtained from the series for 
 $\log Z(h)$ as a function of $h$. Thus the disorder averages of them are  generated from 
 $\rnd{\log Z(h)}$, where  $\rnd{ \ \cdot \ }$ means to take the average over the disorder. 
However, direct calculation of $\rnd{\log Z(h)}$ 
is  formidable challenge in general.   

In order to circumvent the difficulty, $n$ copies (replicas) of the system 
are introduced.  Although they have a common disordered environment, each of
the replica filelds $u^a$ $(a = 1, ..., n)$ couples to independent 
external sources $h^a$. 
Taking the disorder average,  the partition function of the replicated system is defined 
as 
$$
  {\cal Z}(\bh) := \rnd{\prod_{a=1}^n Z(h^a)}= 
  \rnd{\int \prod_{a=1}^n {\cal D} u^a \  
  \rme^{\sum_a \be (-H_{\rm DO}[u^a] + N \, h^{a} \tilde{u}^{a})}}, 
$$
where $\bh := (h^1, ..., h^n)$. 
Employing ${\cal Z}(\bh)$, the generating function per site $\tilde{w}_N (\bh)$ is 
introduced  as
\beq
 \tilde{w}_N (\bh) := \frac{1}{N \beta} \log {\cal Z}(\bh) = \frac{1}{N \beta} \log \rnd{\prod_{a=1}^n Z(h^a)}. 
 \label{def_tildew}
\eeq

Derivatives of $\tilde{w}_N (\bh)$ does not directly yield the thermal cumulants averaged over the disorder.  
It can be transparent using the following notation for the thermal average at $\bh = {\bf 0}$:
$$
  \bra \ \cdot \ \ket := 
  \frac{1}{Z({0})^n}
  \int \prod_{a=1}^n {\cal D} u_a \  \ \cdot \ 
  \rme^{- \sum_a \be H_{\rm DO}[u^a]}. 
$$
Namely, 
$$
  \tilde{w}_N (\bh)  = \frac{1}{N \be} \log \rnd{Z(0)^n \bra \rme^{\sum_a N \be h^a \tilde{u}^a } \ket }. 
$$
One finds that $Z(0)^n$ gives non-trivial effect because it depends on the disorder. 
For instance, the first derivative becomes 
\beq
  \del_a \tilde{w}_N( {\bf 0} ) := \left. \frac{\del  \tilde{w}_N (\bh)}{\del h^a} \right|_{\bh = {\bf 0}} 
  = \frac{\rnd{Z(0)^n \bra \tilde u \ket }}{\rnd{Z(0)^n}}. 
  \label{form_1pt}
\eeq

A usual way of removing the contribution from $Z(0)^n$ is to take 
the zero-replica limit $n \to 0$, which gives
$\del_a \tilde{w}_N({\bf 0}) \to  \rnd{\bra \tilde u\ket}$. 
It apparently seems that  single external source $h$ 
commonly coupled to all the replicated fields is sufficient for generating the disorder 
average of the higher cumulants.  
However, the second derivative at $\bh = {\bf 0}$ becomes 
$$
  \del_a ^2 \tilde{w}_N ({\bf 0})
  \to 
  N \beta  \left( \rnd{\bra {\tilde u}^2  \ket} - \rnd{\bra \tilde u \ket}^2 \right) 
$$
as $n \to 0$. 
The result is slightly different from the desired form. 
For obtaining the correct one, we take the derivative by another source. 
Namely, for $a \neq b$, 
we get 
$$
\del_a \del_b  \tilde{w}_N ({\bf 0})
  \to 
  N \be  \left( \rnd{\bra \tilde u \ket^2}  - \rnd{\bra \tilde u \ket}^2 \right)
$$
as $n \to 0$. Then the second thermal cumulant  averaged over the disorder 
is derived as 
\beq
 N \be \rnd{\bra {\tilde u}^2 \ket - \bra \tilde u\ket^2  } = 
  \lim_{n \to 0} \left( \del_a ^2 \tilde{w}_N ({\bf 0}) - \del_a \del_b  \tilde{w}_N ({\bf 0}) \right). 
   \label{form_2pt} 
\eeq
The above computation demonstrates that 
we need  (at least) two external sources for deriving the disorder average of the 
second thermal cumulant.  It also implies that 
we need at least $p$ replicas coupled with $p$ independent sources for 
the disorder average of the $p$-th thermal cumulant.  This fact clearly indicates    
inconsistency with the zero-replica limit, so that 
we do not use the limit in the present study. 

An alternative way of removing the effect $Z(0)^n$ is to substitute the normalized partition function 
\cite{WL07}
\beq
  z(h) := \frac{Z(h)}{Z(0)}
  \label{def_z}
\eeq
for $Z(h)$. 
Namely,   instead of $\tilde{w}_N(\cdot)$ in (\ref{def_tildew}),  we adopt  $w_N(\cdot)$ defined as the following:
\beq
  w_N(\bh) := \frac{1}{N \be} \log \rnd{ \prod_{a=1}^n z(h^a)} 
  = \frac{1}{N \be} \log \rnd{ \bra \rme^{N \be \sum_{a=1}^n \tilde{u}^ah^a} \ket}.
  \label{def_wN-1}
\eeq
It is normalized in the sense that $w_N ({\bf 0}) = 0$. Computation similar 
to (\ref{form_1pt}) and (\ref{form_2pt}) yields 
\beq
  \del_a w_N({\bf 0}) = \rnd{\bra \tilde{u} \ket}, \ \ 
  (\del_a^2 -  \del_a \del_b) \, w_N({\bf 0}) = N \be \rnd{\bra \tilde{u}^2 \ket - \bra \tilde{u} \ket^2}
  \label{form_12pt}
\eeq
for $a \neq b$. 

Now we take the thermodynamic limit 
$$
  w(\bh) := \lim_{N \to \infty} w_N(\bh), 
$$
and define the effective potential $\gamma (\cdot)$  by the Legendre transform:
\beq
    \gamma (\bvphi) := \sup_{\bh} \(  \sum_{a=1}^n \varphi^a \, h^a -  w (\bh) \). 
    \label{def_ga}
\eeq
The earlier work of Le Doussal and Wiese showed,  with help of the replica trick,  that 
the effective  potential  of a random elastic model defined from the unnormalized generating function
(\ref{def_tildew})  
becomes non-analytic in a glass phase if $\varphi^a = \varphi^b$ for $a \neq b$ \cite{LW02,LW03}.   

Another definition of the effective potential is a free energy with  
 an  order parameter fixed. It is referred to 
 as the constraint effective potential (up to an additive constant) 
 \cite{FK75,OWY86}. 
 Applying this definition to the replicated system, 
we first introduce the density  function $\rho_N(\cdot)$ as 
\beq
 \rho_N (\bvphi) :=  \rnd{\bra \prod_{a=1}^n \de\left(\varphi^a - \tilde{u}^a \right)\ket}. 
 \label{def_rho}
\eeq
The constraint effective potential $\hat\gamma(\bvphi) $ is defined as  
\beq
  \hat\gamma(\bvphi) := 
  - \lim_{N \to \infty} \frac{1}{N \be} \log \rho_N (\bvphi). 
    \label{def_tga}
\eeq
 From (\ref{def_wN-1}) and (\ref{def_rho}), we have 
$$
  \rme^{N \be w_N(\bh)} = \int d \bvphi \, \rho_N (\bvphi) \, \rme^{N \be \sum_a \varphi^a h^a}, 
$$
which implies that $\hat\gamma(\bvphi)$ formally satisfies 
\beq
  w\left(\bh\right) = \sup_{\bvphi} \( \sum_a \varphi^a h^a - \hat\gamma(\bvphi) \). 
  \label{form_w}
\eeq
From (\ref{def_ga}) and (\ref{form_w}), we find that 
 $\gamma(\cdot)$ is the double Legendre transform of 
$\hat\gamma(\cdot)$, which implies that 
$\gamma(\cdot)$ is the convex hull (envelope) of 
$\hat\gamma(\cdot)$ \cite{OWY86}.  

The relationship 
between $\gamma(\cdot)$ and $\hat{\gamma}(\cdot)$ mentioned above 
is nicely explained in the language of the large deviation principle (LDP) \cite[p.23]{T09}.
According to the literature, $\be \hat{\gamma}(\cdot)$ is called a rate function. 
The Legendre transform of it, which is $\be w(\cdot)$ in the present work, 
is called the scaled cumulant generating function. 
The double Legendre transform of the rate function, $\be \ga(\cdot)$,  
is shown to be the convex envelope of $\be \hat{\gamma}(\cdot)$. 

Physical meaning of the effective potential is understood from (\ref{def_tga}).  The probability density 
for the order parameter can be written as 
$$
  \rho_N (\bvphi) \simeq {\rm const. } \, \rme^{- N \be \hat\gamma(\bvphi)}
$$
for large $N$. We see that $\bvphi$ giving minimum of $\hat\gamma(\cdot)$ is realized 
in the thermodynamic limit.  The second thermal cumulant (\ref{form_2pt}) can be 
computed as 
$$
  N \be \int d\bvphi \left( (\varphi^1)^2 - \varphi^1 \varphi^2 \right) \rho_N (\bvphi)
  \simeq {\rm const.} \int d\bvphi \left( (\varphi^1)^2 - \varphi^1 \varphi^2 \right)  
  \rme^{- N \be \hat\gamma(\bvphi)}. 
$$
We see that at least two replicas are needed for the derivation. 

In order to understand relationship between a value of the order parameter and 
form of the effective potential, 
it is instructive to show a mean-field model for the Ising ferromagnet in pure system.  
The order parameter $\varphi$ is the  magnetization per site.  In the high-temperature phase,  the graph of 
$\hat\gamma(\cdot)$ forms like a single well, which has the unique minimum at the origin. 
It leads to the vanishing order parameter.
On the other hand, in the low-temperature phase, the graph of $\hat{\gamma}(\cdot)$ 
forms a double-well potential symmetric under the $Z_2$ transform 
$\varphi \to - \varphi$.  One of the two minima is chosen under a specific boundary condition. 
Thus a value of the order parameter does not vanish in the low-temperature phase.  The  
other effective potential, $\gamma(\cdot)$, is the convex envelope of $\hat\gamma(\cdot)$, 
whose graph has the flat bottom connecting the two minima of $\hat{\gamma}(\cdot)$. 
The consequence $\hat\gamma(\cdot) \neq \gamma(\cdot)$ originates from the mean-filed interaction,
 where arbitrary two spins are interacting. 
If the spin interaction is sufficiently short-ranged, we can show that 
$\hat\gamma(\cdot) = \gamma(\cdot)$ \cite{OWY86}. This is because a value of $\varphi$ can be changed by 
moving a domain wall just adding boundary energy, which vanishes in the thermodynamic limit.

\section{The REM in a magnetic field and its generating function}
\label{sec_REM}
In this section, we first recall the REM and 
derive the generating function for its replicated system. 

The random energy model (REM) is defined on configurations of 
$N$ Ising spins $\bsi:=\{\sigma_1,...,\sigma_N\}$, 
where every $\si_i$ 
 takes the values of $\pm 1$ \cite{D80,D81}. When there is no external field, the energy $E_\bsi$ of 
a spin configuration $\bsi$ is completely independent of how the configuration is. 
It just follows a Gaussian probability density $P(\cdot)$ specifying  disordered  environment:
\beq
  P(E_{\bsi}) := \frac{1}{\sqrt{\pi N J^2}} \exp \left(-\frac{E_\bsi^2}{N J^2} \right). 
  \label{def_P}
\eeq

After  magnetic field $h$ is turned on, 
the energy $E_\bsi$ gets dependence on the magnetization 
$M_\bsi := \sum_{i=1}^N \sigma_i$ and is modified to $E_\bsi - h\, M_\bsi$.  Letting $m_\bsi$ be 
the magnetization per site $M_\bsi/N$, 
the partition function becomes 
\beq
  Z(h) := \sum_\bsi \rme^{-\be E_{\bsi} + \be N m_\bsi h}. 
  \label{def_Z}
\eeq

Now we compute the generating function for the replicated system of the REM.  As we stressed in the 
previous section, we use the normalized 
generating function $w_N(\bh)$ defined by (\ref{def_wN-1})   instead of  
$\tilde{w}_N(\bh)$ plus the replica trick. 
As a by-product, we do not need the free parameter $n$, so that 
we can investigate the simplest but non-trivial case,  $n = 2$. 
Namely, 
 we deal with 
\beq
  w_N(h^1, h^2) := \frac{1}{N \be} \log \rnd{z(h^1) z(h^2)}, 
  \label{def_wN2}
\eeq
where $z( \cdot)$ is defined in  (\ref{def_z}) with use of (\ref{def_Z}). 
It is easily checked in the same way as for (\ref{form_12pt}) that 
$w_N(h^1, h^2)$ actually generates the disorder average of the second cumulants for 
$m_\bsi$ by the following formula:
\beq
  \rnd{\bra m_\bsi^2\ket - \bra m_\bsi \ket^2} = 
   \frac{1}{N \be}  \left(\del_1^2 w_N (0, 0) 
   -   \del_1 \del_2 w_N (0, 0) \right). 
   \label{form_chiN}
\eeq

In order to compute the right-hand side of  (\ref{def_wN2}), 
let us derive  a general formula to $\rnd{\bra O(\bsi^1, \bsi^2) \ket}$, where $O(\bsi^1, \bsi^2)$ 
depends on the replicated spin configurations, but not on the quenched random variables
$\{E_\bsi\}$. The thermal average for the two replicas is 
 $$
  \bra O(\bsi^1, \bsi^2) \ket := \frac{1}{Z(0)^2} \sum_{\bsi^1, \bsi^2} 
  O(\bsi^1, \bsi^2) \rme^{-\be E_{\bsi^1} - \be E_{\bsi^2}}. 
$$
 Since each of $E_\bsi$ independently follows (\ref{def_P}), 
we split the summation into the two cases, $\bsi^1 = \bsi^2$ and $\bsi^1 \neq \bsi^2$, 
 when we take the disorder average. 
Thus we have 
\beqa
\fl
 \rnd{ \bra O(\bsi^1, \bsi^2) \ket} =
 \rnd{ \frac{1}{Z(0)^2}  \rme^{-2 \be E_{\bsi}}} \sum_{\bsi}  O(\bsi, \bsi)
  \nn\\
   + \rnd{ \frac{1}{Z(0)^2}  \rme^{-\be E_{\bsi^1}-\be E_{\bsi^2} } }
  \sum_{\bsi^1\neq \bsi^2} O(\bsi^1, \bsi^2). 
  \label{form_EO}
\eeqa
The first factor is written as 
\beq
  \rnd{ \frac{1}{Z(0)^2}  \rme^{-2 \be E_{\bsi}}} = 
  2^{-N} \rnd{ \frac{1}{Z(0)^2}  \sum_\bsi \rme^{-2 \be E_{\bsi}}} 
  = 2^{-N} \rnd{Y_N}, 
  \label{form_1stFactor}
\eeq
where 
$$
Y_N := \frac{1}{Z(0)^2}  \sum_\bsi \rme^{-2 \be E_{\bsi}}
$$ 
is known as the participation ratio \cite[p.100]{MM09}.  
The second factor is also expressed using $\rnd{Y_N}$ as 
\beq
\fl
   \rnd{ \frac{1}{Z(0)^2}  \rme^{-\be E_{\bsi^1}-\be E_{\bsi^2} } }
   =  \frac{1}{2^N(2^N-1)}\rnd{ \frac{1}{Z(0)^2}  
   \sum_{\bsi^1 \neq \bsi^2}\rme^{-\be E_{\bsi^1}-\be E_{\bsi^2} } }
  = \frac{1 - \rnd{Y_N}}{2^N(2^N-1)}. 
  \label{form_2ndFactor}
\eeq
Insertion of  (\ref{form_1stFactor}) and (\ref{form_2ndFactor}) to (\ref{form_EO}) leads to 
\beq
   \rnd{ \bra O(\bsi^1, \bsi^2) \ket} =
  p_N \, 2^{-N}\sum_{\bsi}  O(\bsi, \bsi)
  + 
  (1-p_N) \, 2^{-2N} \sum_{\bsi^1,  \bsi^2} O(\bsi^1, \bsi^2). 
  \label{form_EO2}
\eeq
Here we have used the notation
$$
  p_N := \frac{\rnd{Y_N} - 2^{-N}}{1-2^{-N}}, 
$$
which obviously has the same thermodynamic limit as $\rnd{Y_N}$. 
 According to the literature \cite[p.101, p.153]{MM09} (see also \cite{G13}), 
\beq
\lim_{N\to \infty} p_N = 
   \lim_{N\to \infty} \rnd{Y_N} = 
  \left\{
    \begin{array}{ll}
      0 & (\be < \be_c) \\
      1-\frac{\be_c}{\be} &  (\be \geq \be_c)
    \end{array}
    \right., 
    \label{res_YN}
\eeq
where $\be_c :=2\sqrt{\log 2}/J$ is the critical temperature 
dividing the paramagnetic phase $(\be < \be_c)$ 
and the glass phase $(\be > \be_c)$ in the REM \cite{D80}.

Now we turn back to (\ref{def_wN2}).  
Using the explicit form of $z(\cdot)$,  we have 
\beq
  \rnd{z(h^1) z(h^2)} = \rnd{\bra \rme^{\be N ( h^1 m_{\bsi^1} + h^2 m_{\bsi^2})}\ket }. 
  \label{form_z1z2}
\eeq
The right-hand side is simply evaluated letting $O(\bsi^1, \bsi^2) = 
\exp(\be N(h^1 m_{\bsi^1}+h^2 m_{\bsi^2}))$ in (\ref{form_EO2}). 
Since the exponent is regarded as a Hamiltonian of non-interacting Ising spins 
in a uniform magnetic field, we find that the right-hand side of (\ref{form_z1z2}) is written as 
\beq
 \rnd{z(h^1) z(h^2)} = A_N+B_N
 \eeq
with 
\beqa 
 A_N &:=&  p_N \left( \ch\left(\be \left(h^1+h^2\right)\right) \right)^N
 \nn\\
 B_N &:=&  (1 -  p_N) \left( \ch(\be h^1) \ch(\be h^2) \right)^N. 
  \label{def_AB}
\eeqa

Let us take the thermodynamic limit of $w_N(h^1, h^2)$. 
\beq
  w(h^1, h^2) := \lim_{N \to \infty} w_N(h^1, h^2) = 
  \lim_{N \to \infty}\frac{1}{N\be} \log(A_N+B_N). 
  \label{def_w}
\eeq

When $\be < \be_c$, $p_N \to 0$ as  $N \to \infty$ according to 
(\ref{res_YN}), so that  $A_N$ vanishes in the thermodynamic limit. Thus we get  
\beq
  w(h^1, h^2) = \frac{1}{\be} \left( \log \ch(\be h^1) +\log \ch(\be h^2) \right). 
  \label{form_highw}
\eeq
Namely $w(h^1, h^2)$  is analytic on the whole $h^1h^2$ plane in the high-temperature phase.

\begin{figure}[h]
\begin{center}
\setlength{\unitlength}{1mm}
		\includegraphics[scale=0.6]{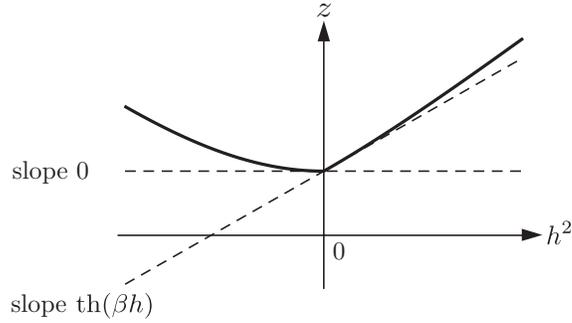}
\end{center}
\caption{
The solid curve is the graph of $z = w(h, h^2)$ with fixed $h>0$. 
The dashed lines represent tangential lines at $h^2=0$.  
The dashed lines with slopes $0$ and $\tah(\be h)$
are respectively the left and the right derivative at $h^2=0$. 
}
\label{fig_wSection1}
\end{figure}

On the other hand, when $\be \geq \be_c$, we need to find which exponentially dominates
 $A_N$ or $B_N$
for large $N$.  It is readily determined if we notice that 
 $\ch (a +b ) = \ch a \ch b + \sh a \sh b$.  The result is 
\beqa
  w(h^1, h^2) = \left\{
  \begin{array}{ll}
  \frac{1}{\be}  \log \ch\left(\be \left( h^1 + h^2\right)\right) & (h^1 h^2 \geq 0)\\[2mm]
  \frac{1}{\be} \left( \log \ch(\be h^1) + \log \ch(\be h^2) \right) & (h^1 h^2<0) \\[2mm]
  \end{array}
  \right..
  \label{form_loww}
\eeqa
It is continuous on the whole $h^1h^2$ plane but not 
differentiable on the lines $h^1=0$ and $h^2 = 0$. In fact,   
\beq
   \del_a w(h^1, h^2) = \left \{
  \begin{array}{ll}
  \tah \left(\be \left( h^1 + h^2\right)\right)  & (h^1 h^2 > 0) \\
 \tah \left(\be  h^a \right) & (h^1 h^2 <0)
  \end{array}
  \right.
  \label{form_phi}
\eeq
for $a=1,2$. It indicates that 
$\del_a w(h^1, h^2)$ is not continuous on $h^a=0$. 
For example, when $h>0$, we get
\beqa
  \lim_{h^2 \uparrow 0} \del_2 w(h,h^2) &=&  0,  
  \nn\\ 
 \lim_{h^2 \downarrow 0} \del_2 w(h, h^2) &=& \tah\(\be h\) \neq 0. 
\label{form_del2w}
\eeqa 
This non-analytic behaviour, which is depicted in Fig.\ref{fig_wSection1}, 
plays an crucial role to differentiability of the effective potential. 

Note that we cannot apply the formula (\ref{form_chiN}) after taking the thermodynamic limit
 since the partial derivatives do not exist on the lines $h^a=0$ $(a=1,2)$. 
For finite $N$,  straightforward calculation gives
\beq
\fl
  \( \del_a^2 - \del_a \del_b \) w_N \(h^1, h^2 \) = \frac{\be  B_N}{c_{a}^2 (A_N+B_N)} 
  + \frac{N \be A_N B_N}{(A_N+B_N)^2}(t_a - t_b) ( t_a-t_{12}).
\label{res_diffwN}
\eeq
where $(a, b) = (1,2)$ or $(a,b) = (2,1)$, and we have used the following abbreviation:
 $c_{a}:= \ch (\be h^a), \  t_a := \tah(\be h^a), \ t_{12} := \tah(\be(h^1+h^2))$.  
Letting $h^1=h^2=0$, we see from  (\ref{form_chiN}), (\ref{def_AB}) and (\ref{res_diffwN}) that 
the susceptibility $\chi_N$ is computed as 
$$
   \chi_N := N \be \rnd{\bra m_\bsi^2\ket - \bra m_\bsi \ket^2}=
   \( \del_a^2 - \del_a \del_b \) w_N \(0, 0\) = \be (1-p_N). 
$$
Using (\ref{res_YN}), we have the thermodynamic limit. 
\beq
 \lim_{N \to \infty} \chi_N = 
  \left\{
    \begin{array}{ll}
      \be & (\be < \be_c) \\
      \be_c &  (\be \geq \be_c)
    \end{array}
    \right.,
    \label{res_chi}
\eeq
 which is precisely equal to the susceptibility first obtained by Derrida \cite{D80,D81}, as expected. 
Note that it holds for both $(a, b) = (1,2)$ and $(a, b) =(2,1)$, which reflects that 
the replica symmetry is preserved in the finite system when $(h^1, h^2) =(0,0)$.  

The same result is obtained by the following limiting procedure with the explicit replica-symmetry breaking 
by $(h^1, h^2) = (h, 0)$.  If $a=1, b=2$, we get 
$$
  \lim_{h \to 0} \lim_{N \to \infty}   \( \del_1^2 - \del_1 \del_2 \) w_N \(h, 0\) = 
  \left\{
    \begin{array}{ll}
      \be & (\be < \be_c) \\
      \be_c &  (\be \geq \be_c)
    \end{array}
    \right.
$$
according to (\ref{res_diffwN}). 
In this formula, however, the replica indices are {\em no longer} exchangeable.   In fact, 
if $a=2, b=1$
$$
    \lim_{h \to 0} \lim_{N \to \infty}  \( \del_2^2 - \del_2 \del_1 \) w_N \(h, 0\) = 
  \left\{
    \begin{array}{ll}
      \be & (\be < \be_c) \\
      \infty &  (\be \geq \be_c)
    \end{array}
    \right.. 
$$
The infinity originates from the second term in (\ref{res_diffwN}) proportional to $N$. 
if $a=1, b=2$, it vanishes because $t_{12} -t_a = 0$ in this term,  while it remains if $a=2, b=1$. 
It can be interpreted as the symmetry by permutation of the replica indices is spontaneously 
broken.  Namely the spontaneous RSB with the original meaning takes place.  
A similar observation is performed in \cite{G13}, where an inter-replica couplings are 
introduced in the Hamiltonian as symmetry breaking terms.

\section{The Effective Potential}
\label{sec_eff}
In this section, we derive the effective potential $\ga(\varphi^1, \varphi^2)$ by the Legendre transform 
(\ref{def_ga}) of $w(h^1, h^2)$. 
Here,  if $w(h^1, h^2)$ is differentiable, the Legendre transform can be carried out 
by solving the following equations 
\beq
  \varphi^a = \del_a w(h^1, h^2), \ \ (a = 1,2 ) 
  \label{form_varphi}
\eeq
for $h^1$ and $h^2$, and then inserting the solutions into the right-hand side of (\ref{def_ga}).  We can easily derive $\ga (\varphi^1, \varphi^2)$ in the 
high-temperature phase along this line.  In fact, from (\ref{form_highw}), 
the equations (\ref{form_varphi}) become 
\beq
  \varphi^a  = \tah \(\be h^a\),  \ \ (a = 1,2 ) 
  \label{form_phiH}
\eeq
for all $h^1$ and $h^2$. Solving them for $h^1$ and $h^2$, we  obtain 
\beq
  \ga(\varphi^1, \varphi^2) = - \frac{1}{\be} \left( s(\varphi^1) + s(\varphi^2) \right),  
  \label{res_gammaH}
\eeq
where 
$$
  s(\varphi) := -\frac{1}{2} \left( (1+\varphi) \log (1+\varphi) 
  + (1-\varphi) \log (1-\varphi) \right). 
$$
It indicates that $\ga(\varphi^1, \varphi^2)$
 has the global minimum at the origin and has no singularity. 

In the low-temperature phase,  first we consider the case of  $h^1 h^2 >0$.  We use the first line of (\ref{form_loww}) for (\ref{form_varphi}),   
which yields
$$
  \varphi^a = \tah\be \left(h^1 + h^2 \right), \qquad (a=1,2). 
$$
It shows that the identity $\varphi^1 =\varphi^2$ holds. Inserting the solution for $h^1+h^2$ to (\ref{def_ga}) yields 
$$
 \varphi^1 h^1 + \varphi^2 h^2 - w (h^1, h^2) =\varphi^1(h^1 + h^2)   - \frac{1}{\be} \log \ch \be (h^1+h^2) 
  = -\frac{1}{\be} s(\varphi^1).  
$$ 
Note that this is the effective potential in the case of $\varphi^1 =\varphi^2$.  

Next we go to the case of $h^1 h^2 < 0$.  Since $w(\cdot, \cdot)$ is given  by the second line of (\ref{form_loww}), 
the result is same as the case of the high-temperature phase (\ref{res_gammaH}).   According to (\ref{form_phiH}), 
the condition $h^1 h^2 < 0$ is translated to $\varphi^1 \varphi^2 < 0$. Thus the results for $h^1 h^2 >0$ and 
for $h^1 h^2< 0$ are summarized as  


\beq
  \ga(\varphi^1, \varphi^2) = \left \{
  \begin{array}{ll}
 - \frac{1}{\be}  s(\varphi^1),    & (\varphi^1 = \varphi^2) \\
  - \frac{1}{\be} \left( s(\varphi^1) + s(\varphi^2) \right)  & (\varphi^1 \varphi^2 <0)
  \end{array}
  \right..
  \label{form_ga}
\eeq
In order to determine  $\ga(\varphi^1, \varphi^2)$ for all $\varphi^1$ and $\varphi^2$
($|\varphi^a| < 1, \, a=1,2)$, 
we have to investigate the 
case of $h^1 h^2=0$.  In this case, a partial derivative does not exist as we have seen in 
the previous section, so that we employ 
  the following geometric interpretation of the Legendre transform (\ref{def_ga}):
for a given $\varphi^1$ and $\varphi^2$,  consider the plane defined by the formula 
\beq
 z = \varphi^1 h^1 + \varphi^2 h^2 +z_0
 \label{def_plane}
 \eeq
in the $h^1 h^2 z$ space. 
We choose $z_0$ in such a way that the plane has a common point with 
the surface $z=w(h^1, h^2)$ and try to minimize the value of $z_0$,   
then the minimum value  equals 
$-\ga(\varphi^1, \varphi^2)$.

\begin{figure}[h]
\begin{center}
\setlength{\unitlength}{1mm}
		\includegraphics[scale=0.6]{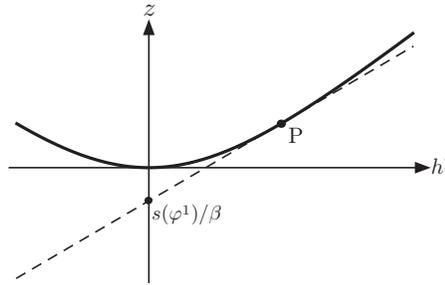}
\end{center}
\caption{The sectional plane $h^2 =0$ in the $h^1h^2z$ space.  
The solid line is the cross section of the surface $z = w(h^1, h^2)$. 
The dashed line represents the plane $z=\varphi^1 h^1 + \varphi^2 h^2 +z_0$ contacting with the surface at $\textrm{P}(h, 0,  w(h, 0))$. 
It intercepts the $z$ axes at 
$s(\varphi^1)/\be$, which is equal to $-\ga(\varphi^1, \varphi^2)$. }
\label{fig_wSection2}
\end{figure}

In order to find the minimum of $z_0$ when $h^1h^2=0$,  
we first consider the case of $h^2=0$ and $h^1 >0$.  Take an arbitrary point $(h, 0)$ with $h>0$ 
and let the corresponding point  on the surface 
$z = w(h^1, h^2)$ be $\textrm{P}(h, 0,  w(h, 0))$.  
We choose $\varphi^1$, $\varphi^2$ and $z_0$ in such a way that 
the plane (\ref{def_plane}) contact with the surface $z=w(h^1, h^2)$ at P.  
Since $\del_1 w(h, 0)$ is well-defined  according to (\ref{form_loww}), 
$\varphi^1$ is uniquely determined as 
\beq
\varphi^1 = \del_1 w(h, 0)  = \tah(\be h). 
\eeq
On the other hand,  $\del_2 w(h, 0)$ does not exist as we have seen in (\ref{form_del2w}). 
In this case, $\varphi^2$ can take the value 
 between the left and the right derivatives, hence  $\varphi^2 \in [0, \tah(\be h)]=[0, \varphi^1]$.  
Since  the point P is on the plane (\ref{def_plane}), 
we find that $z_0 = w(h, 0) - \varphi^1 h = s(\varphi^1) /\be$. 
See Fig \ref{fig_wSection2}. 
Note that if $z_0$ took a value less than $s(\varphi^1) /\be$, the plane 
(\ref{def_plane}) would not have a common point with the surface. 
It indicates that $s(\varphi^1)/\be$ gives the minimum. We thus have
\beq
 \ga\(\varphi^1, \varphi^2 \) = -s(\varphi^1) /\be
 \label{form_ga2}
\eeq
for $\varphi^2 \in [0, \varphi^1]$. 
Similar calculation can be applied in the case when $h^2=0$, $h^1 <0$ 
and we obtain (\ref{form_ga2}) for $\varphi^2 \in [\varphi^1, 0]$. 
%
%
%
%

\begin{figure}[h]
\begin{center}
\setlength{\unitlength}{1mm}
		\includegraphics[scale=0.5]{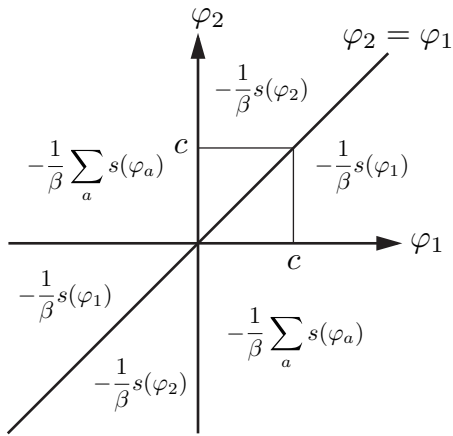}
\end{center}
\caption{Values of $\ga(\varphi^1, \varphi^2)$ on the $\varphi^1 \varphi^2$ plane. 
The segments on $\varphi^1 =c$ and $\varphi^2=c$ show contours with the 
value $\gamma(\varphi^1, \varphi^2)=-\frac{1}{\beta} s(c)$. They meet at $\varphi^1=\varphi^2=c$, where the effective potential becomes non-analytic. }
\label{fig_gamma}
\end{figure}

When $h^1=0$,  exchanging the role of $\varphi^1$ and $\varphi^2$ in 
the case of $h^2=0$, 
we get 
\beq
 \ga\(\varphi^1, \varphi^2 \) = -s(\varphi^2) /\be
 \label{form_ga3}
\eeq
for $\varphi^1\in [0, \varphi^2]$ or $\varphi^1 \in [\varphi^2, 0]$.
Combining the results (\ref{form_ga}) (\ref{form_ga2}) and (\ref{form_ga3}), 
we finally obtain
\beq
  \ga(\varphi^1, \varphi^2) = \left \{
  \begin{array}{ll}
 - \frac{1}{\be}  s(\varphi^1)    &
  (0 \leq \varphi^2 \leq \varphi^1 \ \textrm{or} \  \varphi^1 \leq \varphi^2 \leq 0)\\[2mm]
  - \frac{1}{\be}  s(\varphi^2)   &
  (0 \leq \varphi^1 \leq \varphi^2 \ \textrm{or} \  \varphi^2 \leq \varphi^1 \leq 0)\\[2mm] 
  - \frac{1}{\be} \left(s(\varphi^1) + s(\varphi^2) \right) & (\varphi^1 \varphi^2 <0)
  \end{array}
  \right..
  \label{res_ga}
\eeq
As is shown in Fig.\ref{fig_gamma}, regions that specify the values of 
$\gamma(\varphi^1, \varphi^2)$ have the boudaries $\varphi^a = 0$ 
$(a=1,2)$ and $\varphi^2 = \varphi^1$, on which it is continuous but non-analytic. 
%
%
%
%

The non-analyticity on $\varphi^1 = \varphi^2$ is  observed in fixed-point potentials of 
the Functional renormalization group transformation in various disordered systems having short-range interaction \cite{F86,F01,LW02,LW03,TT04,WL07,LMW08,TT08-2}.  
Following \cite{TT08-2}, it is convenient to introduce the variables  $x:=(\varphi^1 + \varphi^2)/2$ and 
$y:= (\varphi^1 - \varphi^2)/2$.  For fixed $x >0$ and for small $y$ satisfying $|y|<x$, 
the effective potential is written as 
\beq
  \ga(\varphi^1, \varphi^2) = -\frac{1}{\be} s\( x + |y|\). 
\eeq
We see the linear cusp at $y=0$, which resembles the non-analytic effective potential 
in random $O(N)$ models studied in \cite{LW02, LW03,TT08-2}. It should be noted that 
the order of the singularity is  different.  In fact, in the $O(N)$ model, the linear cusp $|y|$ appears in 
the second derivatives of the effective potential with respect to $\varphi^1$ and $\varphi^2$.

\section{The constraint effective potential}
\label{sec_constraint}
In this section we exactly calculate the constraint effective potential for two-replica system of the REM. 
Here the density  function defined in  (\ref{def_rho}) needs slight modification 
 in accordance with discrete spin variables, i.e.,  
$$
 \rho_N(\varphi^1, \varphi^2) :=  \rnd{\bra \de\left(\varphi^1, m_{\bsi^1}\right)\, 
  \de\left(\varphi^2,  m_{\bsi^2}\right)\ket}, 
$$
where $\de(x,y)=1$ if $x=y$ and $\de(x,y)=0$ otherwise. 
The constraint effective potential 
$\hat{\gamma}(\varphi^1, \varphi^2)$ is defined as in (\ref{def_tga}):
$$
  \hat{\gamma}(\varphi^1, \varphi^2) := 
  - \lim_{N \to \infty} \frac{1}{N \beta} \log \rho_N (\varphi^1, \varphi^2). 
$$
In the present study, we easily calculate $\rho_N(\cdot, \cdot)$ 
employing (\ref{form_EO2}). 
\beq
\rho_N(\varphi^1, \varphi^2)  =
   p_N  \, 2^{-N} n(N \varphi^1) \, \de(\varphi^1, \varphi^2) 
   + \left(1-p_N\right) 2^{-2N} \, n(N \varphi^1)\, n(N \varphi^2), 
\eeq
where  
$$
  n(M) := {N \choose \frac{N+M}{2}}
$$
is the number of configurations that have the total magnetization $M$. 
 Employing the Stirring 
formula, we have 
\beq
  \rho_N(\varphi^1, \varphi^2)\simeq 
  p_N \, \rme^{Ns(\varphi^1)} \de(\varphi^1, \varphi^2) + 
  \left(1-p_N\right) \, \rme^{N(s(\varphi^1)+ s(\varphi^2))}
\label{form_rho}
\eeq
for large $N$. 
When $\be \leq \be_c$, since  $p_N \simeq 0$ from (\ref{res_YN}),
we get 
$$
  \hat{\gamma}(\varphi^1, \varphi^2) =  - \frac{1}{\be} \left(
  s(\varphi^1)+ s(\varphi^2)
  \right), 
$$  
which is identical with $\ga(\varphi^1, \varphi^2)$ obtained in (\ref{res_gammaH}).   
When $\be > \be_c$, we find that 
the first term in (\ref{form_rho}) dominates on the line $\varphi^1 = \varphi^2$, thus we conclude that 
\beq
  \hat{\gamma}(\varphi^1, \varphi^2) = \left \{
  \begin{array}{ll}
 - \frac{1}{\be}  s(\varphi^1)    &
  (\varphi^1 =  \varphi^2 )\\[2mm]
  - \frac{1}{\be}  \left( s(\varphi^1) +s(\varphi^2) \right)  &
  (\mbox{otherwise})
  \end{array}
  \right..
  \label{res_tga}
\eeq
The difference between (\ref{res_ga}) and (\ref{res_tga})
 can be understood from a general argument in \sref{sec_EP}.   Namely, $\gamma(\cdot, \cdot)$
is the convex hull (envelope) of $\hat{\gamma}(\cdot, \cdot)$. 

\begin{figure}[h]
  \centering
  \begin{subfigure}{0.3 \columnwidth}
    \centering
    \includegraphics[width = \columnwidth]{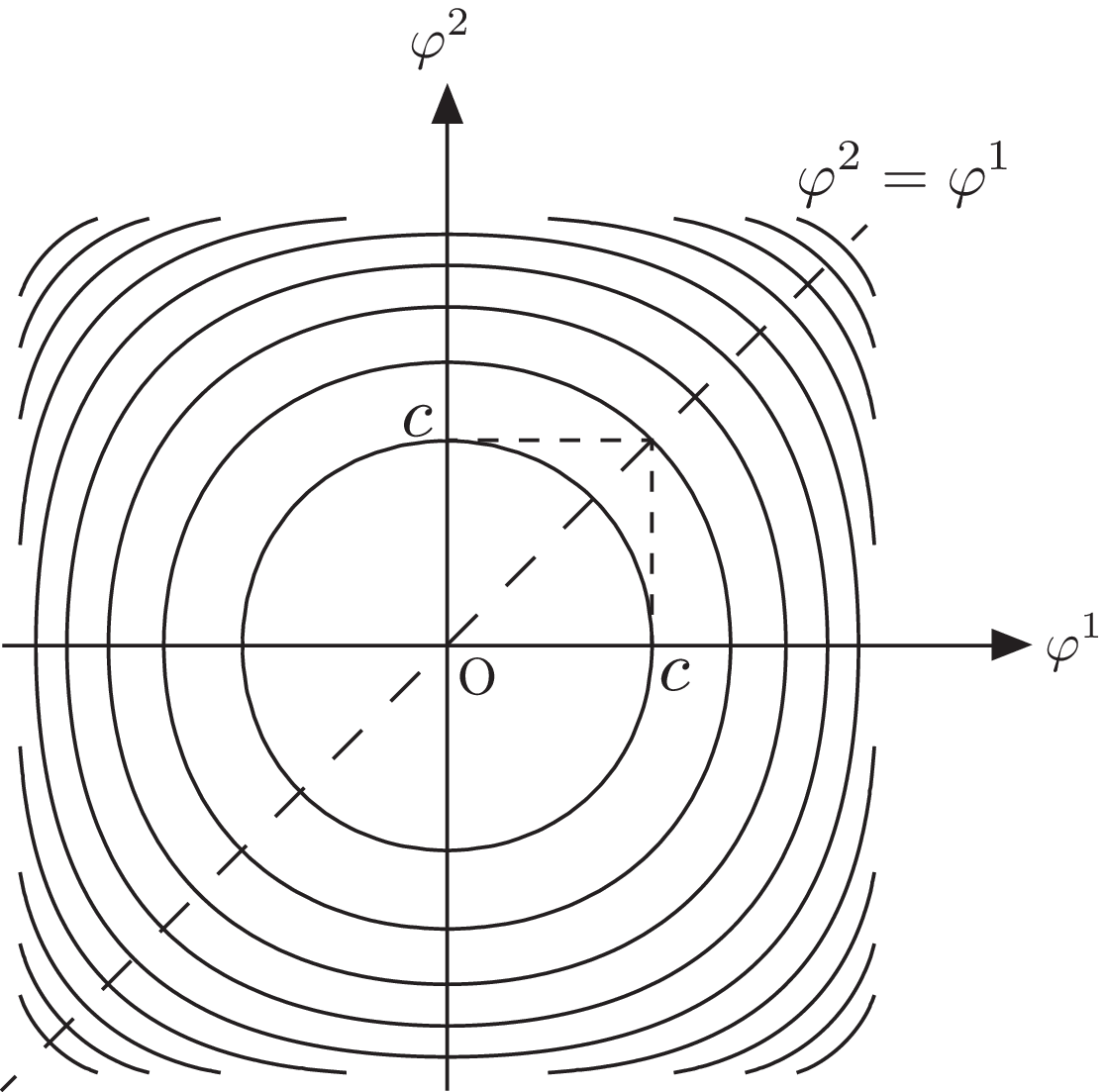}
    \caption{}
    \label{fig:a}
   \end{subfigure}
   \hspace{3mm}
  \begin{subfigure}{0.3 \columnwidth}
    \centering
    \includegraphics[width = \columnwidth]{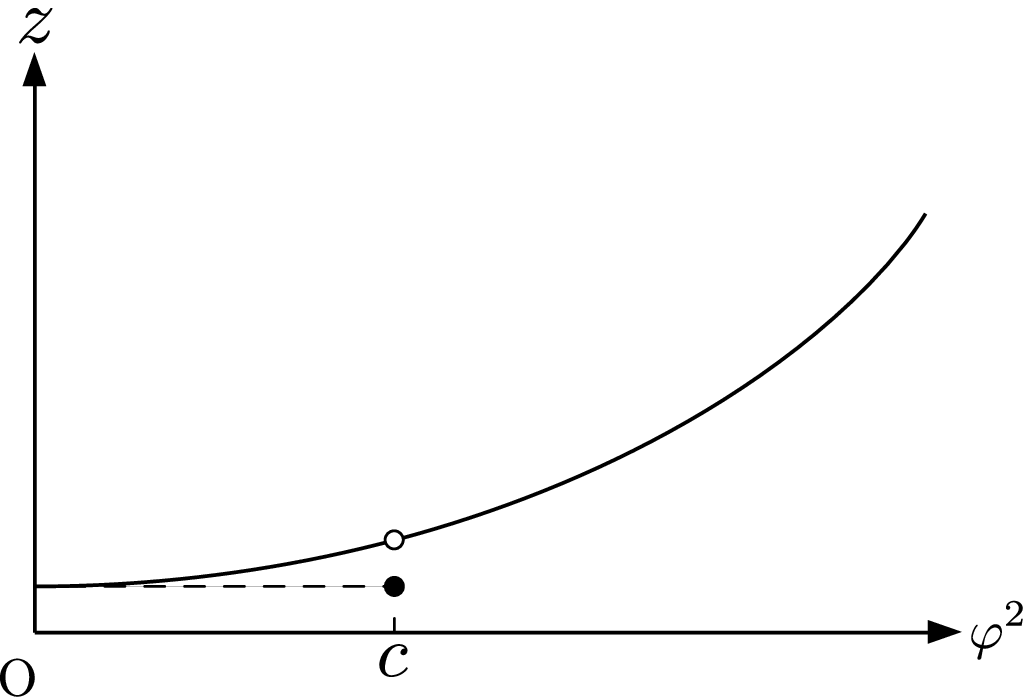}
    \caption{}
    \label{fig:b}
   \end{subfigure}
    \hspace{3mm}
 \begin{subfigure}{0.3 \columnwidth}
    \centering
    \includegraphics[width = \columnwidth]{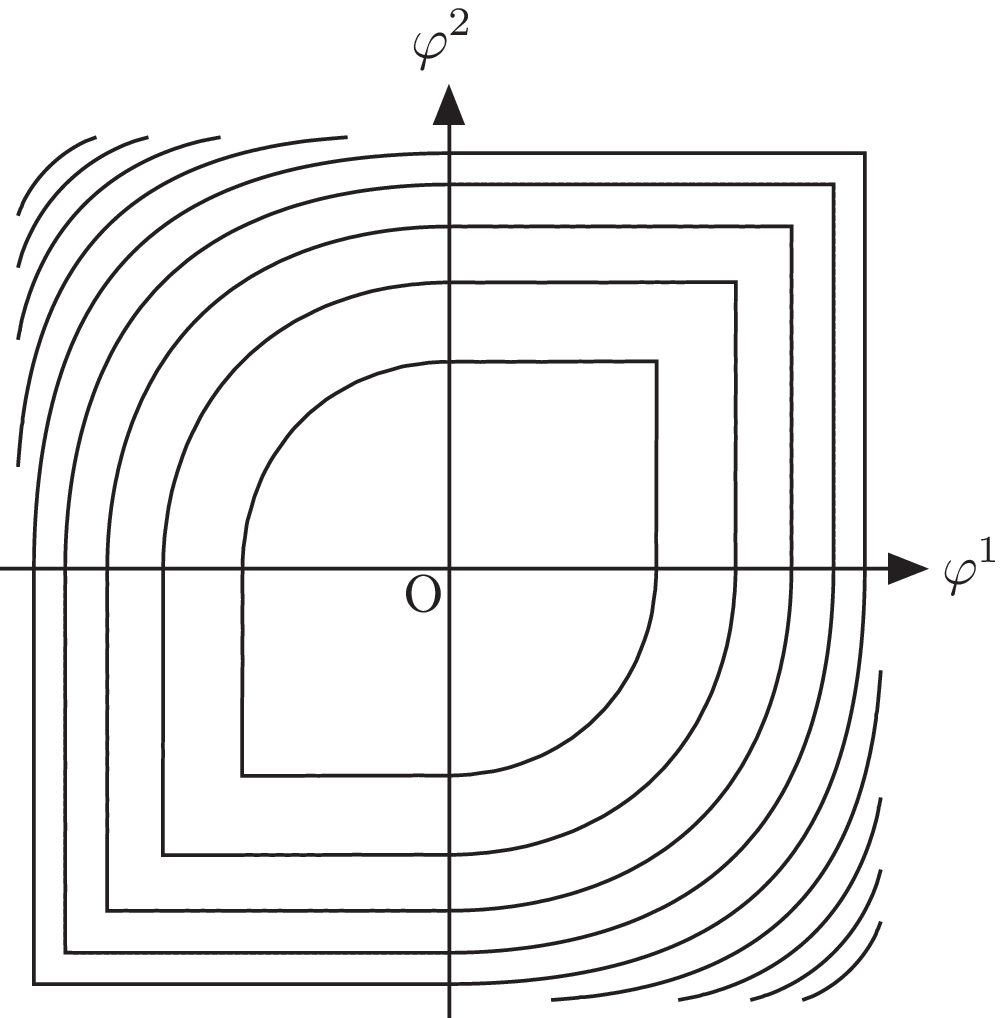}
    \caption{}
    \label{fig:c}
   \end{subfigure}
   \caption{Graphs for $\hat{\ga}(\varphi^1, \varphi^2)$ and for its convex envelope. 
   (a) Contours for the surface $z = \hat{\ga}(\varphi^1, \varphi^2)$. 
   (b) The graph of $z=\hat{\ga}(c, \varphi^2)$ for fixed $c$. 
   (c) Contours for the convex envelope of $z=\hat{\ga}(\varphi^1,\varphi^2)$ }
\end{figure}

This fact can be confirmed by the following argument: \fref{fig:a} shows contours for the graph 
$z= \hat{\ga}(\varphi^1, \varphi^2)$. It has the minimum at the origin. It should be noted  that 
it is discontinuous on $\varphi^2 = \varphi^1$, thus the solid curves are not applicable on the line 
$\varphi^2 = \varphi^1$.  The explicit formula (\ref{res_tga}) indicates that 
$$
  \hat{\ga}(c,c) = \hat{\ga}(0, c) =  \hat{\ga}(c, 0). 
$$
Therefore, the section of the surface by $\varphi^1 = c$  becomes as shown in \fref{fig:b}. 
We also have the same curve for the section by $\varphi^2 =c$.  It implies that we can 
make the convex hull of $\hat{\ga}(\cdot, \cdot)$ by connecting $(c, c, \hat{\ga}(c,c))$ to 
$(c, 0, \hat{\ga}(c,0))$, and to $(0, c, \hat{\ga}(0,c))$ with the horizontal 
segments.  The resultant surface has the contours in \fref{fig:c}. 
It coincides with contours of $\gamma(\cdot, \cdot)$. See \fref{fig_gamma}. 

From the view point of the construction of $\gamma(\cdot, \cdot)$ from $\hat{\ga}(\cdot, \cdot)$, 
we can conclude that the non-analyticity in $\gamma(\cdot, \cdot)$ results from the discontinuity of 
$\hat{\ga}(\cdot, \cdot)$. 

\section{Summary and Discussion}
In this paper,
we have exactly derived the effective potentials of the two-replica system consisting of the REM following the two definitions (\ref{def_ga}) and (\ref{def_tga}). 
It is found that $\ga(\varphi^1, \varphi^2)$, which is defined by (\ref{def_ga}), 
 is continuous but non-analytic on $\varphi^1 = \varphi^2$ 
in the low-temperature phase.  
The result is similar to the effective potential in $O(N)$ models studied in \cite{LW02,LW03,TT08-2}
although the order of the 
singularity is different.  The other effective potential $\hat\gamma(\varphi^1, \varphi^2)$ defined 
by (\ref{def_tga}) 
is discontinuous on the line $\varphi^1=\varphi^2$. The potential surface on this line 
becomes lower than vicinity and has a gap.  Since $\gamma(\cdot, \cdot)$ is the 
convex envelope of $\hat{\gamma}(\cdot, \cdot)$, we can interpret 
that the non-analyticity of $\gamma(\cdot, \cdot)$ is caused by 
 the discontinuity appearing in  $\hat{\gamma}(\cdot, \cdot)$.   
 
In order to see the origin of the discontinuity in detail, 
let us consider the probability density of the replica overlap in the REM
$$
  P_N (q): = \rnd{\bra \de \left(q, \frac{1}{N} \sum_{i=1}^N \si^1_i \si^2_i \right) \ket }. 
$$ 
The right-hand side is evaluated using (\ref{form_EO2}) as 
$$
  P_N (q) \simeq p_N \de \left( q, 1\right) + (1-p_N) e^{N s(q)}
$$
for large $N$. 
In the low-temperature phase, we have the well-known thermodynamic limit 
 (e.g., \cite[p.162]{MM09},\cite[p.180]{B06},  \cite{P03})
 $$
   \lim_{N \to \infty} P_N (q) = \left( 1 - \frac{\be_c}{\be} \right) \de(q, 1) +  \frac{\be_c}{\be} \, \de(q, 0). 
 $$
Thus, when we pick out the two states following the Boltzmann measure, 
the probability for the two states to become identical each other is  non-negligible. 
It happens because a smaller-than-exponential set of configurations dominates 
the Boltzmann measure, which is  referred to as the 
condensation phenomenon \cite[p.100]{MM09}. 
It causes the discontinuous gap 
of the surface $z = \hat{\ga}(\varphi^1, \varphi^2)$ along $\varphi^1=\varphi^2$.  Consequently, 
$\ga(\varphi^1, \varphi^2)$, the double Legendre transform of $\hat{\ga}(\varphi^1, \varphi^2)$,  
becomes non-analytic on $\varphi^1=\varphi^2$.  

Since the condensation phenomenon is considered
 as a typical feature of a glass phase in mean-field models, 
it is plausible that non-analytic effective potential appears together with the RSB as far as 
mean-field models are concerned. 
However, it is unclear whether the same mechanism takes place in short-ranged disordered models. 
 In fact, the non-convexity of $\hat{\ga}(\cdot, \cdot)$ 
will strongly depend on mean-field property of the REM, while 
thermodynamic stability in short-ranged models
ensures convexity of a constraint effective potential \cite{OWY86}.   Further investigation 
will shed light on universal relationship between the non-analyticity and the RSB.

\ack
The author would like to thank G. Tarjus, V. Dotsenko, M. Tissier and 
S. Suzuki for fruitful discussions, comments and criticism.  He is greatly indebted to H. Tasaki for 
interesting discussion, which has considerably  improved this work. 
He is also grateful to LPTMC (Paris 6) for kind hospitality, where most of this work has been done. 

\section*{References}
\bibliographystyle{iopart-num}
\bibliography{myrefs}
\end{document}